\documentclass[useAMS,usenatbib]{mnras}
\usepackage{graphicx} 
\usepackage{natbib}
\usepackage{textgreek}
\usepackage{hyperref}
\usepackage{amsmath}
\usepackage{fixltx2e}
\usepackage{float} 
\usepackage{wrapfig}

\graphicspath{{Pictures/}} 

\title[Brightness variations of MBAs from PanSTARRS-1]{Brightness variation distributions among main belt asteroids from sparse light curve sampling with Pan-STARRS 1}
\author[A. McNeill, A. Fitzsimmons, R. Jedicke et al.]{A. McNeill$^{1}$\thanks{E-mail:
amcneill11@qub.ac.uk},  A. Fitzsimmons$^{1}$, R. Jedicke$^{2}$, R. Wainscoat$^{2}$, L. Denneau$^{2}$, P. Vere\v{s}$^{2,3}$ 
\newauthor E. Magnier$^{2}$, K.C. Chambers$^{2}$, N. Kaiser$^{2}$ and C. Waters$^{2}$ \\
$^{1}$Astrophysics Research Centre, Queen’s University Belfast, BT7 1NN, Northern Ireland, UK.\\
$^{2}$Institute for Astronomy, University of Hawaii at Manoa, Honolulu, HI 96822, USA\\
$^{3}$Department of Astronomy, Physics of the Earth and Meteorology, Comenius University, Mlynsk ́a dolina, Bratislava, 942 48, Slovakia\\}

\begin{document}

\pagerange{\pageref{firstpage}--\pageref{lastpage}} \pubyear{2002}

\maketitle

\label{firstpage}

\begin{abstract} 

The rotational state of asteroids is controlled by various physical mechanisms including collisions, internal damping and the Yarkovsky-O'Keefe-Radzievskii-Paddack (YORP) effect. We have analysed the changes in magnitude between consecutive detections of $\sim$ 60,000 asteroids measured by the PanSTARRS 1 survey during its first 18 months of operations. We have attempted to explain the derived brightness changes physically and through the application of a simple  model. We have found a tendency toward smaller magnitude variations with decreasing diameter for objects of $1<D<8$ km.  Assuming the shape distribution of objects in this size range to be independent of size and composition our model suggests a population with average axial ratios $1:0.85\pm 0.13:0.71\pm 0.13$, with larger objects more likely to have spin axes perpendicular to the orbital plane. 

\end{abstract}

\begin{keywords}
minor planets, asteroids: general - methods: statistical
\end{keywords}

\section{Introduction}

The main asteroid belt situated between Mars and Jupiter contains approximately 95\% of all bodies reported to the Minor Planet Center. The rotational state and evolution of objects in this region are governed by the interplay of several different mechanisms. These include collisional effects and thermal forces such as the Yarkovsky and YORP effects (see below). In this work we investigate the current spin state of main belt asteroids using sparse-lightcurve data obtained by the Pan-STARRS1 (PS1) survey telescope \cite{tonry2012}.

Previous work on asteroid spin statistics using sparse light curve sampling was carried out by \cite{szabo2008} using detection pairs from the Sloan Digital Sky Survey. This investigation derived an approximate shape distribution for a population of $\sim 11,000$ main belt asteroids. Using a similar method with PS1, the aim of our investigation was to look for evidence of YORP reorientation of the rotational spin axes among main belt asteroids.  In lieu of a sample of precise light curves, this large number of detections allowed the comparison of findings from observational data to a statistical model to draw conclusions about the shape and spin pole distributions of these objects.

\subsection{Asteroid Rotation} 

Since their initial formation the rotational behaviour of asteroids in the main belt has undergone considerable evolution. The main factors influencing evolution of rotational behaviour are collisions between asteroids, tidal interactions with large bodies, internal damping and the Yarkovsky-O'Keefe-Radzievskii-Paddack (YORP) effect (\citealt{radzievskii1952}; \citealt{paddack1974}; \citealt{okeefe1975}).

In the absence of external forces, over time the spin state of asteroids will tend toward principal axis rotation, or rotation around the principal axis of the maximum moment of inertia (\citealt{bottke2002}). Objects in an excited state of rotation will lose rotational energy to internal stress-strain cycling and their motion will be damped to principal axis rotation (\citealt{burns1973}). The timescale over which this occurs is given by \cite{harris1994} where $\tau$  \space is the damping timescale in billions of years, $P$ is the rotation period of the asteroid in hours, $D$ is the diameter in kilometres and $C$ is a constant $\simeq 36$, uncertain to within a factor of 2.5 (\citealt{breiter2012}, \citealt{pravec2014}).

\begin{equation} \tau_{damp}\approx\frac{P^3}{C^3D^2}\end{equation}

This shows that larger objects will return to principal axis rotation over shorter timescales than smaller objects. 

\subsection{Collisions} 

Collisions between asteroids can result in the changing of their semi-major axes and spin poles. Major collisions may result in the catastrophic disruption of one or both of the objects. Collisional effects have played a large role in the evolution of the shapes, sizes and cratering of asteroids in the main belt. The existence of asteroid families with similar orbital and compositional characteristics are a direct result of collisions between large objects causing their catastrophic disruption (\citealt{nesvorny2002}). Sub-catastrophic collisions may also induce non-principal axis rotation causing objects to 'tumble'. For small objects these collisions may happen with greater frequency than the timescale required to damp this motion back to principal axis rotation. Sub-catastrophic collisions could therefore be a driving mechanism producing small tumblers. Such collisions may also serve to alter the spin axis of a body over an average timescale as shown in (\citealt{farinella1998}).

\begin{equation}
\tau_{rot}=\frac{1}{P_iR^2N(>D_{rot})}
\end{equation}

\begin{equation}
N(>D_{rot}) \simeq 1.36\times 10^{6} (D_{rot})^{-2.5}
\end{equation}

\begin{equation}
D_{rot} = (\frac{\sqrt{2}\rho\omega}{5\rho_pv})^{1/3} D_{t}^{4/3}
\end{equation}

Here $P_i$ is the intrinsic collision probability as described by \cite{wetherill1967} with an average value in the asteroid belt of 2.85 x 10\textsuperscript{-18} km\textsuperscript{-2} yr\textsuperscript{-1} (\citealt{farinella1998} ; \citealt{marchi2014}), $D_{rot}$ is the diameter in kilometres of the projectile required to completely change the spin axis of an object of diameter $D_{t}$ (also in kilometres), $\rho$ and $\rho_p$ are the bulk densities of the target and projectile respectively and $v$ is the average collision velocity in kilometres per second. 

The index of -2.5 in equation 3 represents the size distribution power law exponent from the relationship $N(>D)\propto D^{-b}$ and represents an approximation for the size range present in the dataset. The constant in this equation is obtained from the estimate that there are $1.36\times 10^{6}$ objects with $D>1$km. (\citealt{jedicke2002}; \citealt{bottke2005})

\subsection{Thermal Forces}

The Yarkovsky-O'Keefe-Radzievskii-Paddack (YORP) effect causes the spin rate of asteroids and meteoroids to increase or decrease due to thermal torques. For simplicity consider the asteroid as a blackbody radiator with sunlight falling on the surface of the object being absorbed. This will then be emitted as thermal radiation at a direction normal to the surface providing a radiative force normal to the surface. For a spherical or highly symmetrical object these forces will balance giving no net change to the spin rate (\citealt{rubincam2000}). The anisotropic emission from an asymmetric object gives a net torque acting to cause an increase or decrease in the spin rate of the object. For further information, a full review is presented in \cite{bottke2006}.  Physical modelling of the YORP effect has been studied in great detail, however, it is only in recent years that it has been measured directly (\citealt{lowry2007}; \citealt{durech2008}; \citealt{durech2012}; \citealt{lowry2014}; \citealt{rozitis2014}).

The change in spin states due to YORP is the accepted explanation for the difference in rotational frequency distributions between large ($D>40$km) and small asteroids ($D<40$km). For large asteroids the distribution of their rotation rates approximates a Maxwellian distribution, with a period cut-off at $2.2$h (\citealt{pravec2002}).This spin barrier represents the approximate spin rate that would be required for the centrifugal force to overcome the self-gravity of the rubble pile and cause the aggregate to break apart.  The existence of this ‘spin barrier’ is taken as evidence that these objects are ‘rubble piles’ made up of much smaller segments held together by self-gravity and weak cohesive forces rather than single coherent monoliths. In the small body population ($D < 200$ m) the distribution is non-Maxwellian and there has been shown to be an abundance of fast and slow rotators (\citealt{pravec2000}; \citealt{pravec2002}). YORP spin-up to rotation rates beyond this 'spin barrier' is a potential driving mechanism for rotational disruption of asteroids.

YORP will also produce a force acting at an angle to the plane of rotation and hence act on the spin axis orientation (obliquity). If an object spins down, non-principal axis rotation can evolve which will persist until the rotation of the object is damped back to principal axis rotation. It has been observed in simulations (\citealt{vokrouhlicky2002}) that an object affected by YORP will tend toward an asymptotic obliquity value or 'end state' at which it is stable, the body will then remain at this obliquity until its spin pole alignment is further affected by outside forces i.e. collisional effects . In populations for which the YORP timescales are significantly shorter than collisional axis resetting timescales, it would be expected that a significant number of the objects would have spin axis angles clustered around such end state values (\citealt{vokrouhlicky2002}, \citealt{vokrouhlicky2007}). The timescale over which the spin axis of an asteroid will be driven to an asymptotic state by YORP has been given by \cite{rozitis2013} where the YORP timescale $\tau$ in years can be calculated from the YORP rotational acceleration $|\frac{d\omega}{dt}|$ in radians per year; $a$ and $D$ are expressed in AU and kilometres respectively.

\begin{equation}
\tau_{YORP} \approx \omega/|\frac{d\omega}{dt}|
\end{equation}

\begin{equation}
|\frac{d\omega}{dt}| = 1.20_{-0.86}^{+1.66} \times 10^{-2} (a^{2}\sqrt{1-e^{2}}D^{2})^{-1}
\end{equation}

\begin{figure}
  \begin{center}
\includegraphics[width=0.5\textwidth]{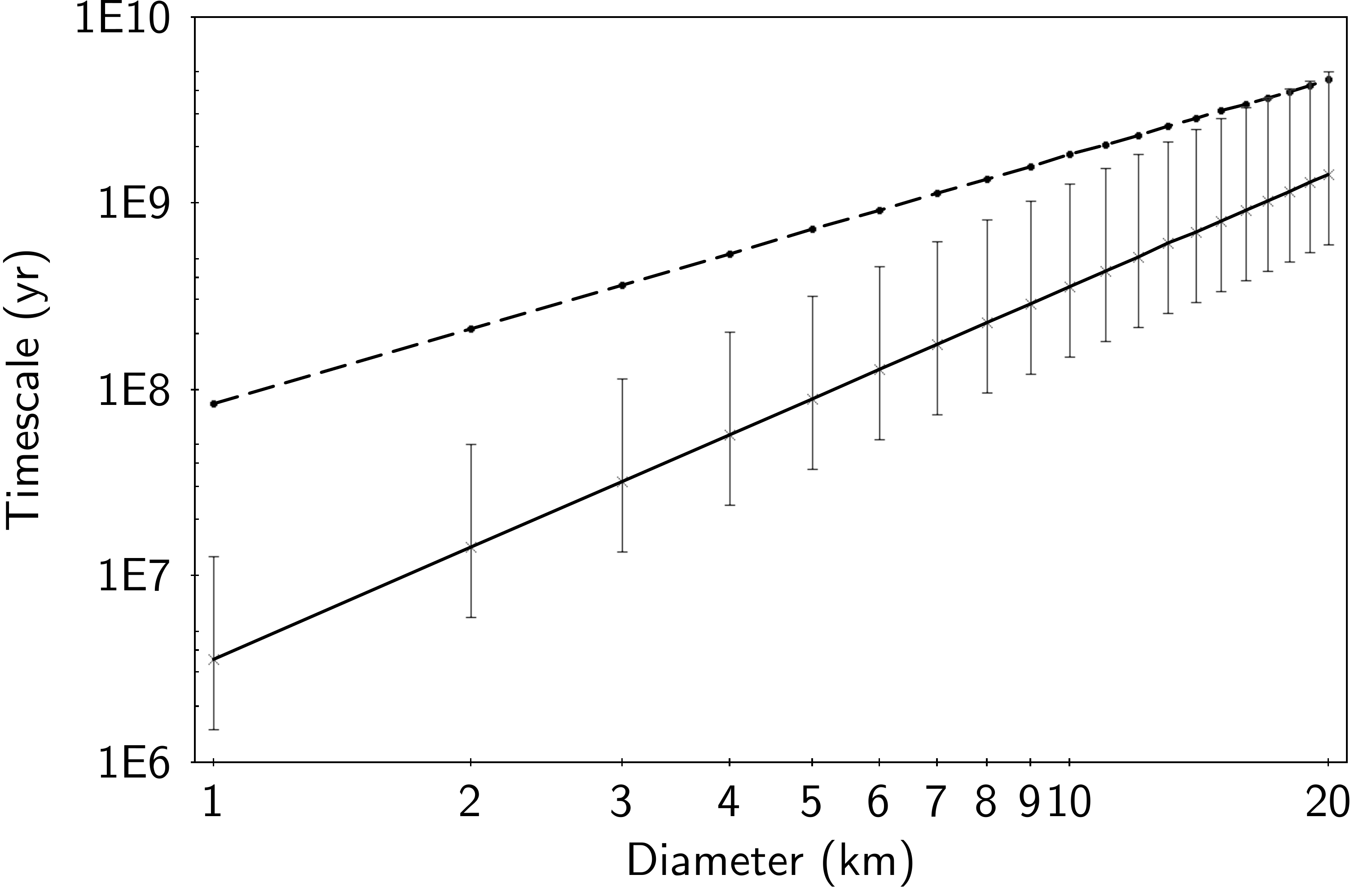}
 \caption{A comparative plot of the timescales for axis resetting by collisions and the YORP effect in the inner main belt ($2.0\leq a \leq 2.5$ AU) determined from Equations 2 and 5. The dashed line represents the collisional axis resetting timescale and the solid line is the median YORP axis resetting timescale with the error bars indicating the range of possible values within 1\textsigma spread.}
\label{innertimescale}
  \end{center}
 \end{figure}

\begin{figure}
  \begin{center}
\includegraphics[width=0.5\textwidth]{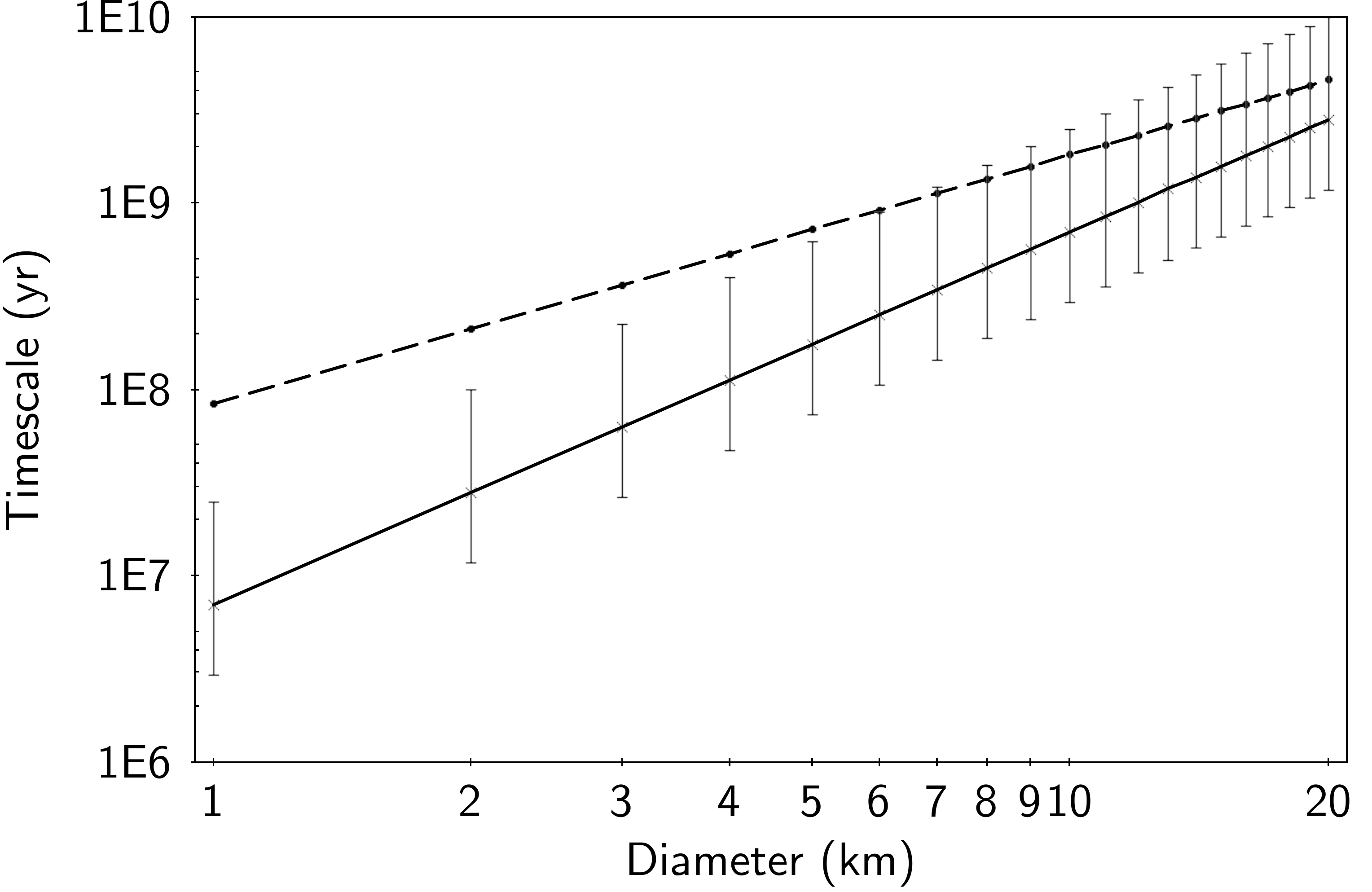}   
\caption{The same as Figure~\ref{innertimescale} except this time for the outer main belt ($3.0\leq a \leq 3.5$ AU). }
\label{outertimescale}
  \end{center}
 \end{figure}

This mechanism will be unable to fully take place in environments in which collisional axis resetting will dominate. Thus a comparison has been made between the spin axis resetting timescale due to to collisions and the axis resetting timescale due to YORP. Figure~\ref{innertimescale} shows that for small asteroids in the inner belt ($2\leq a \leq 2.5$ AU) YORP should dominate the rotational evolution and spin-axis reorientation. Figure~\ref{outertimescale} shows the same calculation for outer main belt asteroids ($3\le a \le 3.5$ AU). Recently, \cite{cibulkova2014} have modelled collisional probabilities for separate regions in the main belt. We found no significant difference in our conclusion when we took this variation with $a$ into account.

\section[]{Pan-STARRS Data}
\subsection{Data Selection}
The data used in this investigation was obtained by PS1 in the first 18 months of survey time.  The 1.8m PS1 telescope is situated on Haleakala on the Hawaiian island of Maui and covers 7 square degrees on the sky. It is equipped with a 1.4 billion pixel CCD camera, currently the largest digital camera of its kind in the world (\citealt{tonry2004}). The system uses six filters as defined in \cite{tonry2012} with the data used in this work taken from the wide w-band ($\sim 400-700$nm) Solar System Survey. The system operates by taking a 45 second exposure of an area of sky and then returning to the same area after approximately 15 minutes. These images are detrended and calibrated
via the Image Processing Pipeline (IPP; \citealt{Magnier2013}). The IPP also subtracts consecutive pairs of images, detects objects remaining in these difference images and passes those detections to the Moving Object Processing System (MOPS; \citealt{denneau2013}). MOPS attempts to link detections of transient objects into 'tracklets' containing the same moving objects, and associate them with previously discovered Solar system bodies. The four consecutive visits per night for each chunk of sky enables discovery of moving objects down to magnitude $m_w\simeq 22$.

During the timespan of the initial survey, PS1 made approximately 1.5 million confirmed detections of moving objects, to which we applied a series of constraints. Only known main belt asteroids were included by filtering according to semi major axis and eccentricity. Additionally, only detections with a formal magnitude uncertainty of $\leq 0.02$  and at
a phase angle $\leq 10^{\circ}$ were included
were considered. This should ensure that magnitudes were not significantly affected by photometric uncertainties or phase-angle effects. After these constraints had been applied a sample of $\sim 60,000$ asteroids with $\sim 264,000$ detections remained.

\subsection{Data Processing}

By determining the difference in apparent magnitude between consecutive detections of an asteroid, we calculated the absolute rate of change of magnitude with respect to time for each object. As the YORP effect will vary according to semi-major axis and albedo, we split the data into smaller datasets based on semi-major axis $a$. The two ranges studied were those asteroids with $2 \le a \le 2.5$ AU (henceforth referred to as the inner belt) and those with $3 \le a \le 3.5$ AU (the outer belt). 

These data sets were further subdivided into 1km diameter bins according to absolute magnitude. This was performed according to equation 7 where $D$ is expressed in kilometres, $H$ is the absolute magnitude of the body and $p_v$ is its albedo.

\begin{equation}
{D} = 2 \times 10^{0.2(29.14-2.5\log p_V - H)}
\end{equation}

The relative abundances of different asteroid spectral types within the main belt is not constant with respect to semi-major axis and as such these ranges were selected due to their relative concentration of one type of asteroid (in this case S-type for the inner belt and C for the outer). Rather than simply taking the average albedo value for S and C-types,
we calculated the mean albedo in the semi-major axis ranges according to the Sloan Digital Sky Survey moving object catalogue (\citealt{hasselmann2012}; \citealt{carvano2010}) for asteroids of diameter greater than 5km. This size limitation was imposed as it represents the lowest diameter at which the taxonomy data within the SDSS is complete (\citealt{demeo2013}).  The mean albedos for asteroids in the two distance ranges were found to be $0.207\pm 0.020$ and $0.103\pm 0.012$ for the inner and outer main belt respectively.

The absolute rates of change in brightness were calculated in units of magnitudes per 15 minutes for each of these diameter bins. The resulting cumulative frequency distributions are plotted in Figures 3 and 4. It is worth noting that the number of asteroids in each bin for $D > 6$km declines sharply, particularly for the inner belt. This is a result of larger asteroids reaching the PS1 saturation limit of $V \approx 15$.

\section[]{Statistical Model}

In order to draw any conclusions from the observed data set it was necessary to create a  statistical model with which to compare it. The model used here generates a synthetic population of ellipsoidal asteroids with assumed shapes and spin pole orientations. $A$ is the apparent cross-section of an asteroid with principal axes, $a$, $b$ and $c$ as seen from Earth. When 
$\theta$ is the angle between the spin axis of an asteroid and the plane of the sky and $\phi$ is the rotational phase, $A$ is given by Equation 8 adapted from \citet{leconte2011}.

\begin{equation}
A=\pi \sqrt{c^{2}sin^{2}\theta(a^{2}sin^{2}\phi + b^{2}cos^{2}\phi)+a^{2}b^{2}cos^{2}\theta}
\end{equation}

The model uses a uniform spin frequency distribution from 1-10.9 day\textsuperscript{-1} across all applicable size ranges, corresponding to rotational periods from of the spin barrier at 2.2h to a period of 24h. This assumption is reasonable when compared with the flat distribution of measured rotational frequencies at small sizes \citep{pravec2002}. The synthetic population is sampled such that two detections separated by 900 seconds, corresponding to the average interval between PS1 images, are generated for each asteroid, with the apparent area of the object at each of these two points in its rotation giving the change in magnitude, $\Delta m$. 

\begin{equation}
\Delta m = \lvert 2.5\log \frac{A_1}{A_2}\rvert=\lvert m_1-m_2\rvert
\end{equation}

For both $m_1$ and $m_2$ a uniformly distributed uncertainty value is selected between -0.02 and 0.02, consistent with the uncertainty values in the selected Pan-STARRS data set. It was also assumed that the change in magnitude between detections was purely due to geometric effects i.e. limb scattering effects were not accounted for.

Setting $a$ equal to $1$ as we are only dealing with magnitude differences, the values of $b$ and $c$ can be varied using various mathematical functions, such as Gaussian or Lorentzian distributions, to approximate the true shape distribution of main belt asteroids in the size range in question. The obliquity $\theta$ is varied according to an arccosine distribution ($acos(x)$ where x is a random number uniformly varied between 0 and 1) for each asteroid between $0^{\circ}$ and a maximum angle limit.

For each population in turn a cumulative distribution function is generated from the randomly sampled brightness variation of each object, and this curve is compared to the observed data using the two sample Kolmogorov-Smirnov test. In order to perform this we need the most complete diameter range available from our data set which is for $2<D<3$ km. The best fit from this size range was then applied to each of the other size ranges in turn, with obliquity remaining a free parameter. This assumes that there will be no significant dependence of shape distribution with size. However, the obliquity distribution of asteroids was not assumed to remain constant.  Using this method the parameters yielding the best fit for the data-set can be determined. 

\section[]{Results}

\subsection[]{Observed Cumulative Distribution Functions for Magnitude Variations}

\begin{figure}
  \begin{center}
\includegraphics[width=0.5\textwidth]{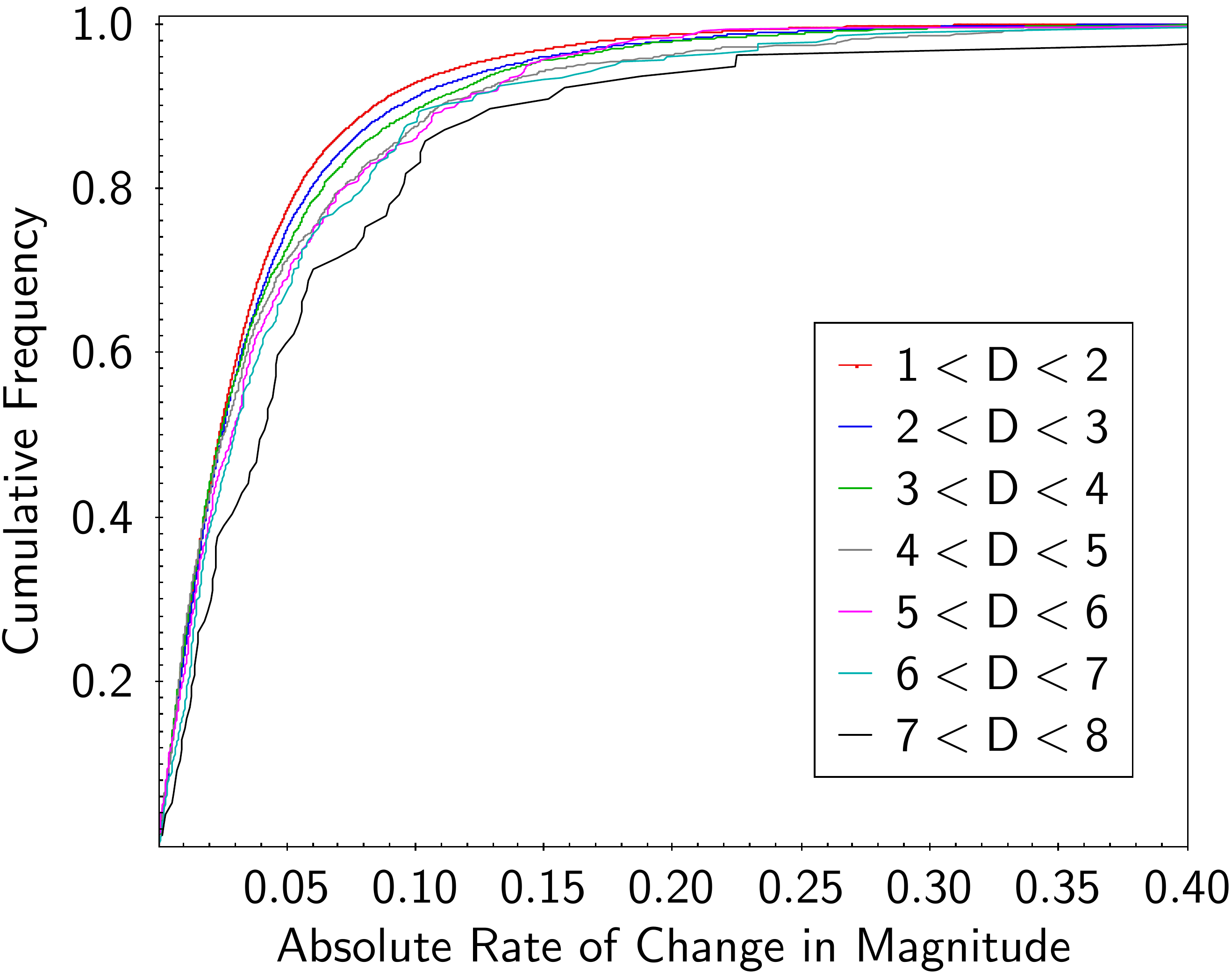}
   \caption{Cumulative frequency curves of the absolute rate of change in magnitude in 1 kilometre diameter bins in the inner main belt.}
\label{innercumul}
  \end{center}
 \end{figure}

\begin{figure}
  \begin{center}
\includegraphics[width=0.5\textwidth]{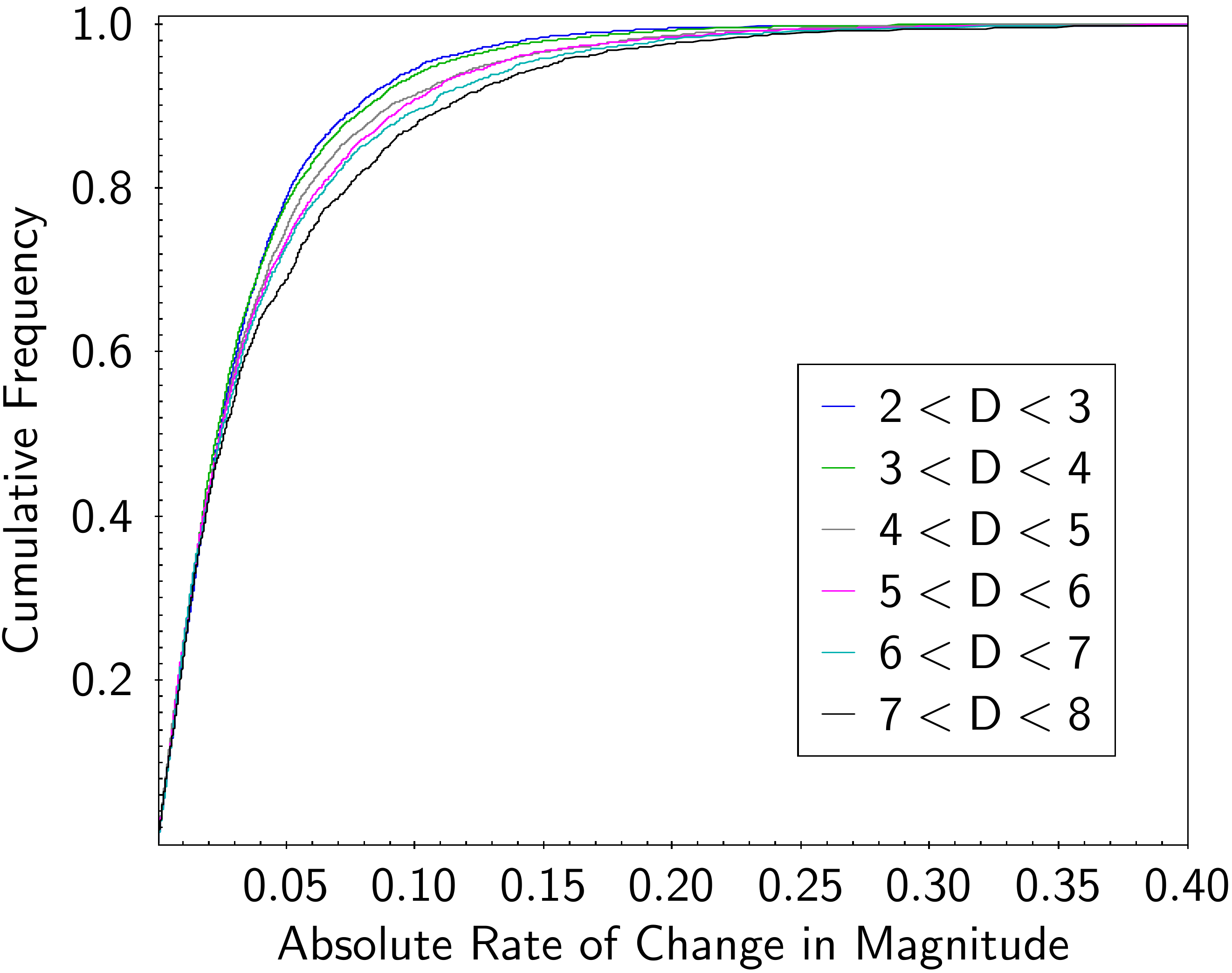}
   \caption{Cumulative frequency curves of the absolute rate of change in magnitude in 1 kilometre diameter bins in the outer main belt.}
\label{outercumul}
  \end{center}
 \end{figure}

The cumulative distribution functions (CDFs) of magnitude variation for asteroids in both the inner and outer regions of the main belt are shown in Figures~\ref{innercumul} and~\ref{outercumul}. For both regions the magnitude variations tend toward larger values with increasing diameter for objects of $1<D<8$ km. Beyond this size the CDF appeared to be relatively stable up to $D\simeq 15$ km. It is worth noting that there were few objects in these higher diameter bins due to the brightness limit of PS1, with virtually no asteroids with diameters $> 15$km accurately measured. Although the overall trend for $D\leq 8$km is the same in both the inner and outer belt, the curves themselves are not identical at each size range. Explanations for this trend  will be explored in section 5.1. 

We considered the possibility that this effect could be due to some systematic bias in the detections. For objects with at least 4 detections within our selection criteria we looked separately at the change between the first two detections in a tracklet and the last two. A similar test was used to verify that there were no systematic errors in SDSS detections in \citet{szabo2004}. There was no significant difference between the two cumulative distribution functions produced using each of these data subsets and the same trend with change in diameter was observed in both cases.

We looked into the possibility that the difference in semi-major axes of the objects within each of our considered regions (inner and outer belt). In order to do this we considered a single size range of objects in the inner belt, in this case the $2<D<3$ km range. This subset of objects was sorted by semi-major axis into bins of width 0.1 AU and a cumulative distribution function constructed, as before, for each bin. There was no statistically significant difference between the CDFs produced in each case.

We also considered the possibility that the difference in proper motions of objects at different semi-major axes could introduce biases into our results. For example, if a slow moving asteroid is in close proximity to a background star or galaxy in the first detection it is probable that this will still be the case in one or more subsequent detections. This may cause problems in background sky subtraction, however, the pair-wise subtraction method used by PS1's Image Processing Pipeline (IPP) should ensure that any significant biases will be minimised. At the opposite end of the scale, if an object is moving quickly in a single exposure then that may lead to trailing in the image. We consider asteroid 79512 from our data set, at a semi major axis of 2.05 AU. This object at the time of its detections was moving with proper motion 0.526 deg/day. An object with this proper motion will produce $< 1$ arcsec trailing when observed with PS1. This is negligible considering the average seeing at the site of PanSTARRS 1 of around 1 arcsec. We are content that the difference in proper motion between objects at different distances is not introducing any significant uncertainties into our data.

Although the data sample used in this investigation was restricted to phase angle $\alpha < 10^{\circ}$. most asteroids exhibit an opposition effect of a sharp non-linear increase in brightness (\citealt{belskaya2000}) at small phase angles. To check if this effect had any bearing on the observed size dependence of variability we divided our data into two sub-samples where $\alpha < 2^{\circ}$ and $8 < \alpha < 10^{\circ}$. In each case the CDFs for the same diameter bins were statistically indistinguishable from each other and the full sample.

The possibility of a bias caused by the ecliptic latitude of the asteroids in our sample was also checked. Our sample was divided into two groups according to ecliptic latitude, $\beta$, where $|\beta| < 20^{\circ}$ and  $|\beta| > 40^{\circ}$.Again there was no statistically significant difference from our earlier result.

Finally, two sub-samples were constructed corresponding to the first 200 days of our data, and the last 200 days of our data, to guard against an unrecognised time variation in the data fidelity. Again, there was no difference between these sub-samples and our original analysis.

\subsection[]{Shape Modelling}

A series of different shape distributions and assumptions were used in an effort to obtain the best fit possible. The axis ratios, $b/a$, of the modelled population were varied in the form of both truncated Lorentzian and Gaussian distributions. Several relationships between b and c were also assumed. These took the form of both rigid relationships e.g. $b=c$ and Lorentzian or Gaussian distributions in $b$ and $c$.

It was initially assumed that all asteroids are prolate spheroids (a\textgreater b=c) with fixed spin pole latitude of $50^{\circ}$, the average value of objects with known spin axes. The value of $b/a$ was varied as a truncated Gaussian of centre $\mu$ and standard deviation $\sigma$ in order to obtain the best possible fit to the observed CDF curves generated from the observational data (Figures~\ref{innercumul} and~\ref{outercumul}). The Gaussian distribution was truncated at the point where $b/a=1$. In this case the best fit was obtained for a median shape of $1:0.83\pm 0.12:0.83\pm 0.12$ from a shape distribution with parameters $\mu = 0.90, \sigma=0.16$.

\begin{figure}
  \begin{center}
\includegraphics[width=0.5\textwidth]{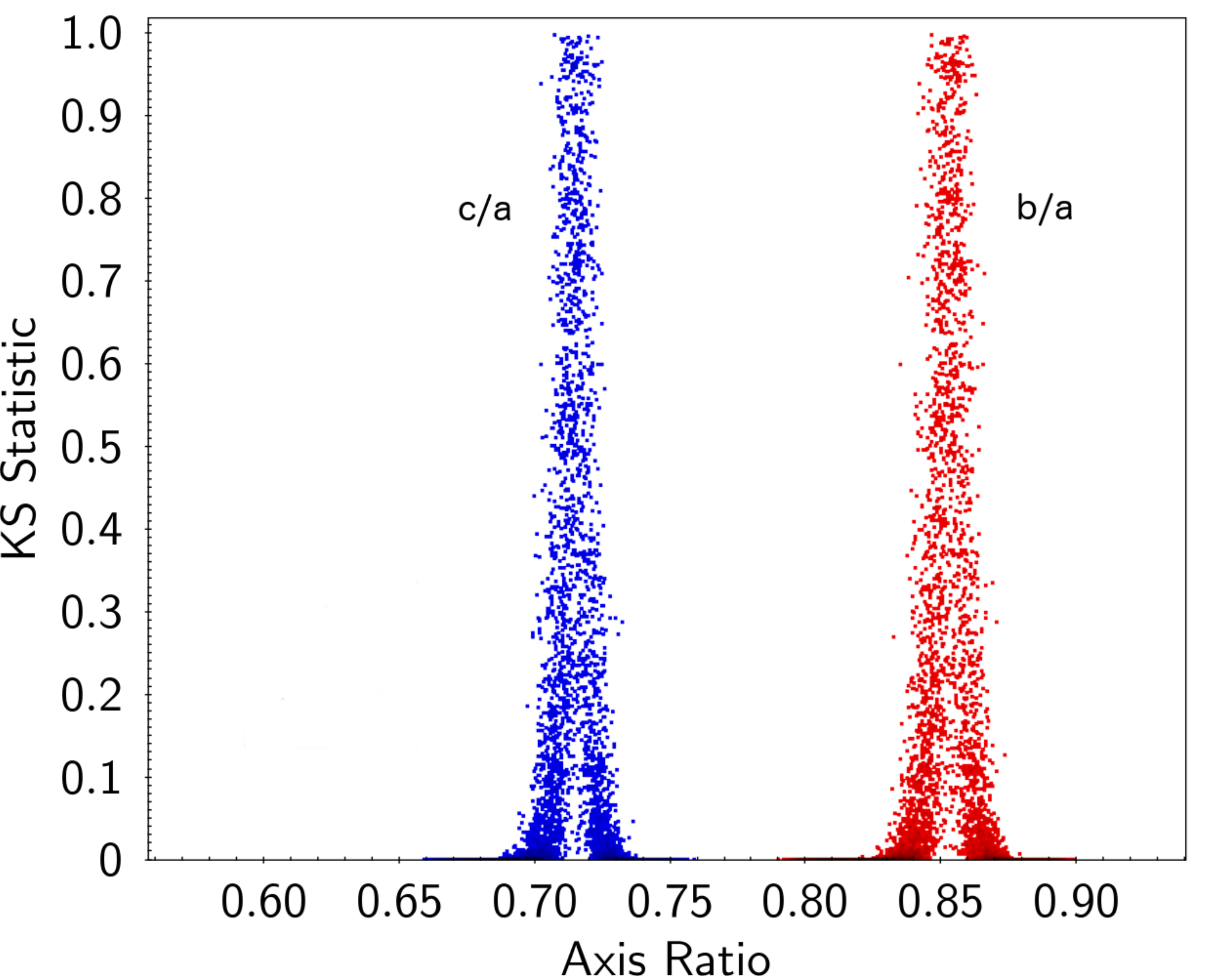}
   \caption{KS statistics from model asteroid populations compared to our dataset as a function of axis ratio. Each of the 5000 points show the KS statistic calculated by comparing the observed CDF to a model population with truncated Gaussian shape distributions. The axis ratio value corresponds to the median shape for each model population.}
\label{averageinner}
  \end{center}
 \end{figure}

To explore other shape distributions, we used the stated values of $b/c$ from the asteroid spin vector and shape data set from \cite{krys2007}. An approximate distribution for these values was determined and applied to the model. Varying $b/a$ produced a best fit for a median shape of $1:0.85\pm 0.13:0.71\pm 0.13$ from a truncated Gaussian distribution of $\mu=0.94, \sigma=0.19$. This was repeated using the median value of $b/c=1.2$ from the database,  no significant difference in the best fit parameters was found. Figure \ref{averageinner} shows the K-S statistic obtained for the mean axis ratios $b/a$ and $c/a$ from 10,000 modelled Gaussian shape distributions. 

\section[]{Discussion}

\subsection[]{Explaining the CDF Trend}

As shown above, the cumulative distribution functions for both the inner and outer main belt show a trend toward smaller variations with decreasing size for asteroids with $D<8$ km. We had assumed that smaller asteroids would be more readily aligned by the YORP effect to spin axes perpendicular to the orbital plane, as the YORP timescale is proportional to the size of the object. This would imply that larger variations in magnitude would be observed at smaller sizes due to the greater alignment of these objects giving higher amplitude measured light curves, the opposite trend to the one we observe. There are several possible explanations for this unexpected result.

\subsubsection[]{YORP Reorientation}

Assuming the shape distribution is independent of size, the observed size variation in the CDF curves could imply that the spin poles have a tendency to become more aligned to the sky plane with increasing size. It has been observed for small main belt asteroids ($D < 30$km) that the YORP effect on asteroids drives them to asymptotic obliquity values clustered around angles perpendicular to the orbital plane as predicted by theory (\citealt{capek2004}). These asteroids in the main belt will give higher amplitude light curves. Hence when a YORP oriented population is sampled a greater proportion of large magnitude variations will be measured than for a randomly aligned population. Therefore, if small diameter asteroids are more likely to be found at random obliquities and larger objects are more likely to be found at YORP end-states aligned closer to the sky plane,  this would produce a similar trend to the observed CDF curves. 

For this to be the case, larger objects would have to be more reoriented by YORP than smaller objects. This is not expected. Figures 1 and 2 show that the timescales for reorientation by YORP and collisions converge with increasing diameter,  collisional axis resetting becoming the dominant mechanism at $D\geq 20$km. This  suggests that statistically more asteroids with small diameters will be reoriented by the YORP effect. For reorientation to explain the trend in the absolute rates of change in magnitude with size, YORP would have to be acting in tandem with another mechanism.

The Yarkovsky effect (\citealt{bottke2006}) will act most strongly upon small objects and those with shorter semi-major axes. If small diameter asteroids were reoriented to YORP end-states and were then removed from the observed population by Yarkovsky drift, this could explain the inconsistency between the YORP timescale calculations and the observed effect. However, \cite{farinella1999} calculated the average semi-major axis drift of an object due to the Yarkovsky effect as a function of diameter. Their results suggest that the average displacement over a collisional lifetime will be of order $10^{-2}$ AU for asteroids with $1<D<10$ km. Therefore the effect of Yarkovsky removal in the observed spin distribution should be negligible.

\subsubsection[]{Rotation rate dependence on size}

Objects with a faster rotation rate would result in a greater number of large magnitude variations when sampled. Therefore, if at smaller diameters, the rotation rate was slower than for larger asteroids, less variation would be observed for smaller asteroids. This would result in a trend among the CDF curves like that observed here. However there is no evidence in previous observational work of a significant difference in rotation rate with diameter within the size range in question (\citealt{pravec2002}). Figure 5 shows known rotational frequencies from the Light Curve Database (\citealt{warner2009}). The geometric mean of the rotational frequency in this size regime suggests a small decrease in mean rotation frequency with increasing diameter. This behaviour is also observed in recent lightcurve data obtained by the Palomar Transient Factory (\citealt{palomar}). This decrease in rotation rate is the opposite of the behaviour needed to explain our observed trend in increasing magnitude variation with increasing diameter.

\begin{figure}
  \begin{center}
\includegraphics[width=0.5\textwidth]{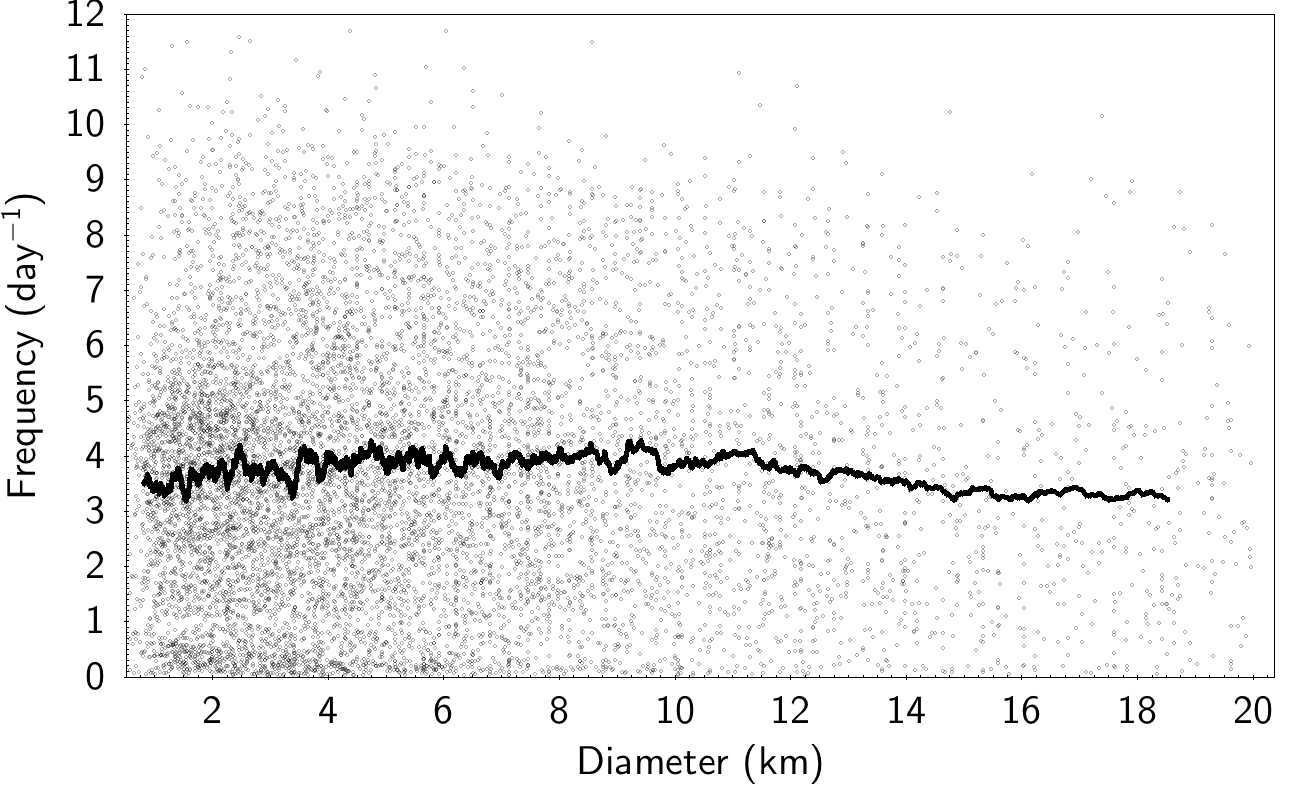}
   \caption{Rotational frequency as a function of diameter for 2647 main belt asteroids stored in the Light Curve Database. The black line represents the running box geometric mean of the catalogued frequencies.}
\label{frequency}
  \end{center}
 \end{figure}

\subsubsection[]{Asteroid elongation dependence on size}

A highly elongated asteroid will produce a high amplitude light curve while a spherical asteroid will display no variation in brightness during its rotation. Therefore if objects were found to be on average more spherical with decreasing size this would result in these asteroids showing smaller variations in magnitude when sampled. This could then explain the trend observed in Figures~\ref{innercumul} and~\ref{outercumul}.

\begin{figure}
  \begin{center}
\includegraphics[width=0.5\textwidth]{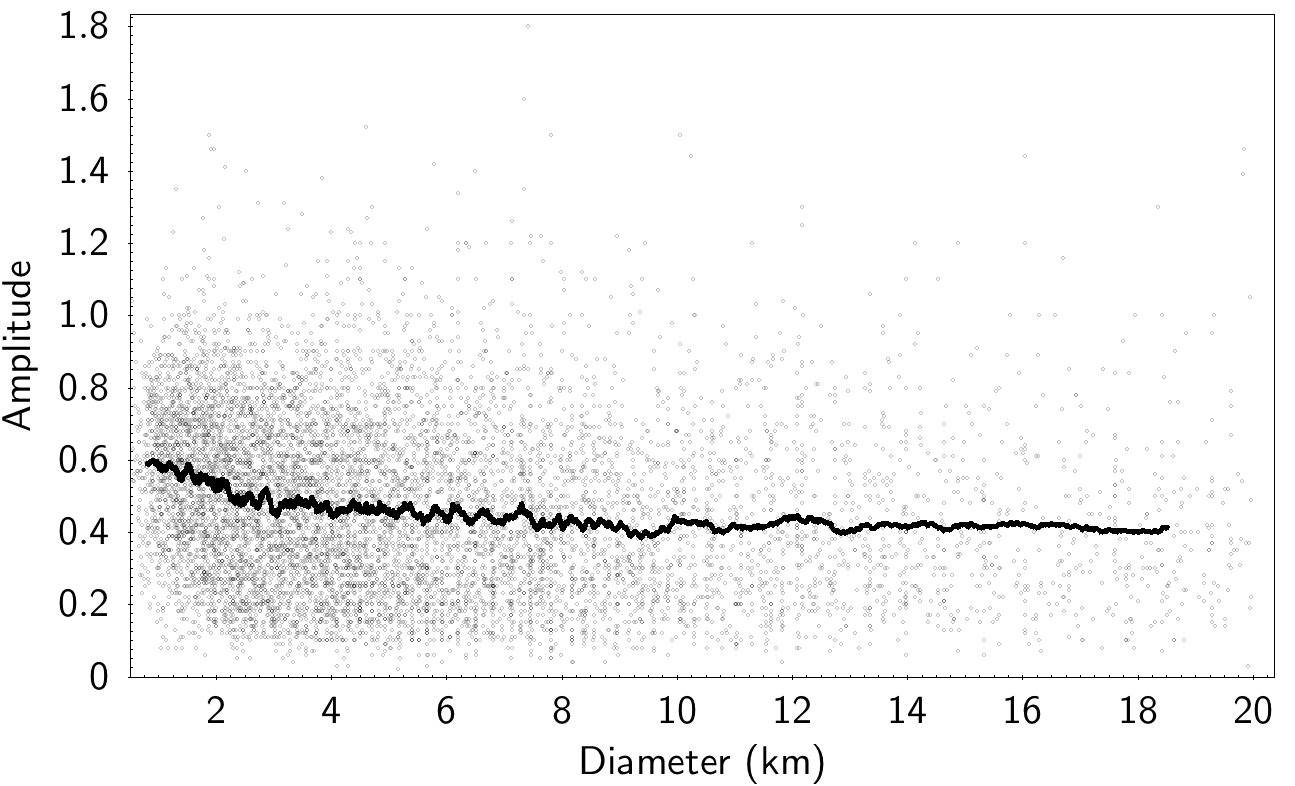}
   \caption{Amplitude as a function of diameter for the same sample of objects as Figure~\ref{frequency}. The black line represents the running box geometric mean of reported amplitudes.}
\label{amplitude}
  \end{center}
 \end{figure}

Figure~\ref{amplitude} shows the measured amplitudes from the Light Curve Database (\citealt{warner2009}) with increasing diameter within the size range contained in our data. Without spin axis information it is not possible to calculate accurate elongation values for each of these objects, however there is no observed trend of increasing amplitude with increasing diameter. This data set will be subject to observational bias as it is much easier to obtain light curves for smaller objects at high amplitudes than low amplitudes. It is possible that a debiased data set would show a greater proportion of low amplitude objects than are seen here and the actual trend may differ. Recent data from the Palomar Transient Factory (\citealt{palomar}) suggests a general trend toward decreasing amplitude with increasing diameter in objects of $D < 100$ km. However, in our size range the geometric mean of their amplitudes appears to be roughly constant.

Both \cite{domokos2009} and \cite{henych2015} simulated the effect of subcatastrophic collisions on the elongation of small asteroids ($D<20$km). They demonstrated that the cumulative effect of collisions should lead to an increase in the target object's elongation, occurring over shorter timescales at smaller sizes. However, \cite{henych2015} state that the estimated timescales for this process to occur are significantly longer than the collisional disruption timescales for the asteroids in question. Therefore, any trend in elongation with size is unlikely to be due to this mechanism.
These observational and theoretical studies support our assumption that elongation should not increase with size for objects in our data. 

\subsubsection[]{Tumbling asteroids}

The presence of tumbling asteroids in the data set could provide a mechanism by which reorientation could occur more frequently at larger sizes. Modelling work on the effect of YORP on tumbling asteroids has shown that the effect will still affect the rotation state of the object as it tumbles, however YORP alone will not easily return the object to principal axis rotation (\citealt{vok2015}). This suggests that the timescale for this process will be longer than the damping of a tumbling object as it returns to rotation around the principal axis of the maximum moment of inertial given in Equation 1. As tumbling asteroids will rarely present the same surface area toward the Sun they can be considered effectively 'immune' to axis reorientation by YORP as the effect will act randomly upon the object as it rotates. The thermal forces never act consistently to drive the object toward a particular obliquity value until the spin state has been damped to near-principal axis rotation.

\begin{figure}
  \begin{center}
\includegraphics[width=0.5\textwidth]{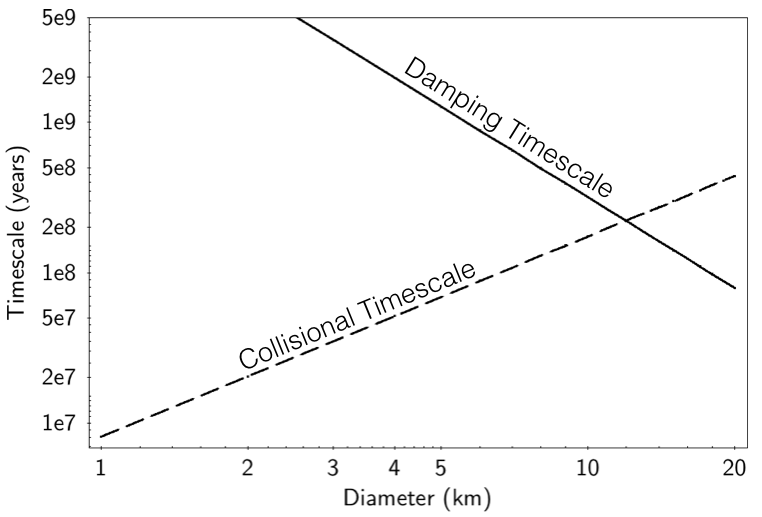}
   \caption{A comparison of the collisional axis resetting timescale for objects with the damping timescale required to bring a tumbling asteroid to principal axis rotation. The dashed line represents the collisional axis resetting timescale obtained from Equations 2-4 for an object with the median period of known tumbling objects in the LCDB (\citealt{warner2009}) with $1<D<10$ km, P=114h . The solid line represents the tumbling damping timescale of the same object calculated using Equation 1.}
\label{tumbling}
  \end{center}
 \end{figure}

A comparison between the timescale over which a typical tumbling asteroid will be damped and its collisional axis resetting timescale is shown in Figure~\ref{tumbling}. At small diameters with $P=114$ hours, the median period of known tumbling objects obtained from the Light Curve Database, the damping timescale is significantly longer than the timescale for axis resetting by collisions.

Therefore if sub-catastrophic collisions act to alter the spin axis orientation of the asteroids and induce non-principal axis rotation, then it follows that small asteroids are likely to undergo another similar collision within the damping timescale. This would suggest that some asteroids at these sizes may be kept constantly tumbling, giving a higher abundance of tumbling asteroids at these size ranges. As there are fewer tumbling asteroids as diameter increases due to fewer collisions and shorter damping timescales, then larger objects are therefore more likely to undergo YORP reorientation hence giving a greater proportion of higher magnitude variations. 

\begin{figure}
  \begin{center}
\includegraphics[width=0.5\textwidth]{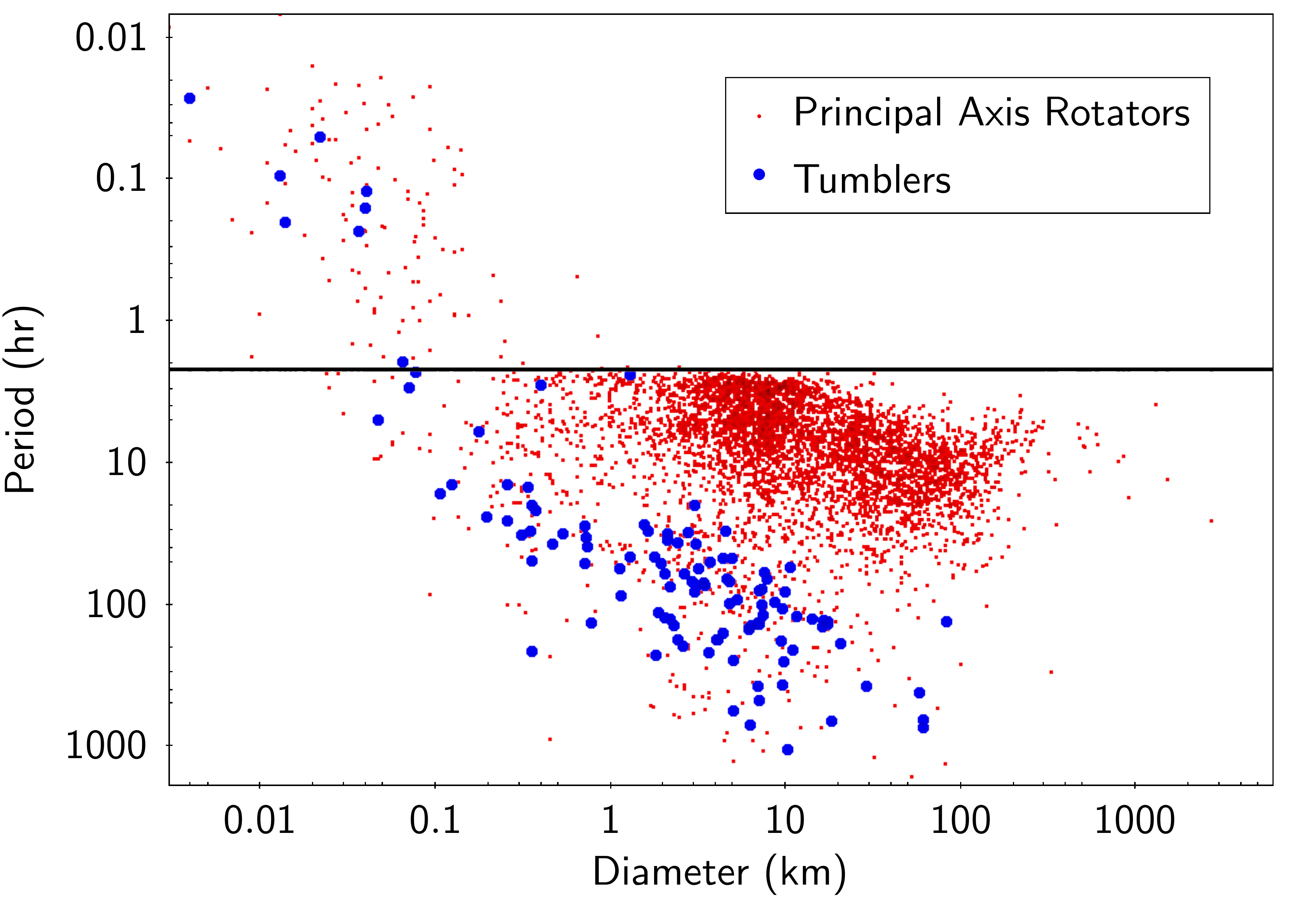}
   \caption{A plot of period against diameter for each of the $\sim$ 5500 asteroids contained in the LCDB. The bold black line represents the spin barrier at 2.2 hours.}
\label{pvsd}
  \end{center}
 \end{figure}

\begin{figure}
  \begin{center}
\includegraphics[width=0.5\textwidth]{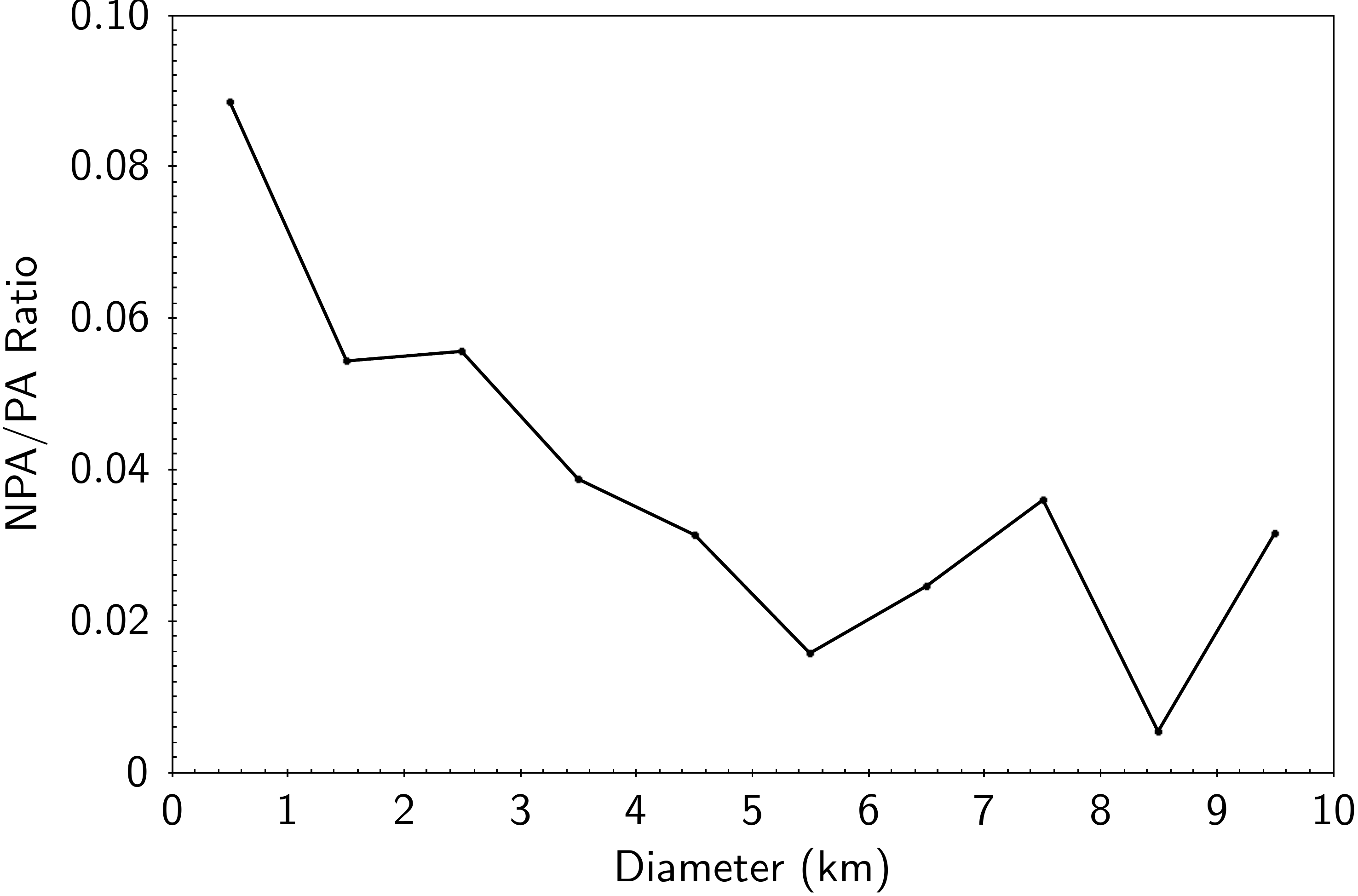}
   \caption{The ratio of non-principal axis rotators to principal axis rotators in 1km diameter bins as listed by the Light Curve Database (\citealt{warner2009})}
\label{ratiovsD}
  \end{center}
 \end{figure}

To look for evidence of this, we again took the Light Curve Database (\citealt{warner2009}). The rotational periods of both principal axis and non-principal axis as a function of size is shown in Figure~\ref{pvsd}. The observed ratio of tumbling asteroids to principal axis rotators in 1km diameter bins was found to decrease with increasing size, and is shown in Figure~\ref{ratiovsD}. This suggests a greater proportion of tumblers at small sizes, and hence a greater proportion of objects prevented from undergoing YORP reorientation. 

To test this we approximated sparse sampling of tumbling objects by introducing a subset of objects with randomly aligned spin axes into our model population. If we assume that all objects in our largest diameter range have spin axes aligned perpendicular to the orbital plane and introduce a population of these pseudo-tumblers we can estimate the proportion of tumbling objects required to explain the difference between our smallest and largest diameter CDFs.

A ratio of pseudo-tumblers to principal axis rotators of $2:3$ was required to give a similar change to that shown in the CDFs. This is an improbably large fraction  when compared to the ratio measured from the Light Curve Database of $\sim 6\%$. Although this value will be subject to a selection bias as it is easier to obtain unambiguous light curves for principal axis rotators, it is unlikely that a debiased  value would reach the $40\%$  that our modelling suggests is required to explain our data. Therefore we conclude that none of the above mechanisms can currently explain our observed trends with size.

\subsection[]{Comparing results for the inner and outer main belt populations}

Figures 3 and 4 show that the rate at which the distribution of magnitude variation changes with diameter is greater for the inner belt than the outer belt. This may be due to either a difference in the reorientation of the objects between the two populations, or a difference in shape distribution. Unlike the difference in variability with diameter, the difference with semi-major axis is potentially more straightforward. Reorientation due to YORP will happen more slowly at greater heliocentric distances and thus objects in the outer belt should show systematically less variability, all other factors being equal. This agrees with what we see here.

To test this we assumed that the shape distribution of the outer belt was identical to that obtained for the inner belt. In order to assess the validity of this assumption, a best fit truncated Gaussian shape distribution was obtained for the outer main belt independently at $2<D<3$ km using the previously outlined method. This gave an average axial ratio of $1 : 0.87\pm 0.09 : 0.74 \pm 0.11$ corresponding to a population with a shape distribution $\mu = 0.90, \sigma = 0.11$. This is not significantly different from that found for the inner belt and therefore our assumption holds. The difference between the two populations must therefore be due to a difference in the degree of reorientation. Part of this may be due to the slower rate of YORP in the outer belt.
At present our statistical model does not account for these effects and this may offer an avenue for future work.

\subsection[]{Previous constraints on shape and spin}

A previous study by \cite{szabo2008} used a similar sparse light curve sampling method, utilising data from the SDSS Moving Object Catalogue. Assuming all objects to have spin axis parallel to the sky plane, they reported a shape distribution in good agreement with axial ratios from then-published light curves. We compared our findings to an updated distribution of axial ratios from current light curves (\citealt{krys2007}). The light curves given in this database suggest an average asteroid shape with 1 \textsigma  \space spread with axial ratio $1:0.76\pm 0.13:0.62\pm 0.11$. This result is within 1 \textsigma \space of our own findings, albeit both \textless $b/a$\textgreater \ and \textless$c/b$\textgreater \ are smaller than  our results.

We then used the truncated Gaussian distribution for $b/a$ and the approximated distribution for $b/c$ obtained from the \cite{krys2007} data set, and varied the  upper obliquity limit in an attempt to obtain a good fit from this large diameter shape distribution. It was possible to obtain fits for object populations in 1km diameter bins down to $\simeq 3$km but not at smaller sizes. This would suggest that the shape distribution for large asteroids cannot be taken to apply to the small size regime. However it is not possible to further investigate this without a full shape analysis of small diameter asteroids. 

\cite{szabo2008} determined a shape distribution for their population by comparing the cumulative frequency curve obtained from the observed detection pairs to a linear combination of template curves of known shapes using $a/b$ from 1.1 to 4.0. This method effectively sets a condition that $b\leq 0.91a$ for any shape distribution obtained through it. This will give a shape distribution with a dearth of near spherical objects, which may explain the more elongated result reported by them compared to our own result.

It is not stated in \cite{szabo2008} which relationship between $b$ and $c$ was used so we assumed that the asteroids are treated as prolate spheroids. We found that it made no significant difference whether $c/b$ was varied as a distribution according to \cite{krys2007} or if $c/b$ was kept constant, therefore we assumed a constant $c=0.8b$. These smaller c axes will affect the shape distribution of the overall population as it will require the asteroids to have slightly larger b axes in order to produce the same brightness variation. Overall this will produce a population with larger average b axes than would be obtained using the assumption that all of these asteroids are prolate spheroids. The latter assumption could explain the smaller average b axis size reported in \cite{szabo2008} compared with our result.

A combination of these differences in methodology may explain the more spherical shapes obtained from our model when compared to \cite{szabo2008}. The more spherical asteroid shapes from our results compared to those in the database of measured light curves found in \cite{krys2007} could be a result of observational biases i.e. the more spherical asteroids our model suggests are present in the MOPS database would result in low-amplitude light curves and thus could be under-represented in reported data.

\section{Conclusions}

Using sparse light curve sampling methods, cumulative distribution functions of the brightness variations of main belt asteroids found by the Pan-STARRS 1 survey were generated. Comparison of the CDF plots for objects sorted into 1km diameter bins shows that for asteroids where $D<8$km there is a trend toward smaller magnitude variations as size decreases. We have considered several possible explanations for this, however, we cannot reproduce these observations at present. 

Using a statistical model it was possible to generate synthetic CDF plots from input shape and spin-axis parameters. It was found that the best fit was obtained for a population of triaxial ellipsoids with average axial ratio $1:0.85\pm 0.13:0.71\pm 0.13$ with smaller asteroids less likely to be aligned with spin axes parallel to the sky plane by the YORP effect. This average shape is more spherical but consistent with the average shape determined from the data found in \cite{krys2007}.

\section*{Acknowledgments}

The Pan-STARRS Surveys have been made possible through contributions of the Institute for Astronomy, the University of Hawaii, the Pan-STARRS Project Office, the Max-Planck Society and its participating institutes, the Max Planck Institute for Astronomy, Heidelberg and the Max Planck Institute for Extraterrestrial Physics, Garching, The Johns Hopkins University, Durham University, the University of Edinburgh, Queen’s University Belfast, the Harvard-Smithsonian Center for Astrophysics, and the Las Cumbres Observatory Global Telescope Network, Incorporated, the National Central University of Taiwan, and the National Aeronautics and Space Administration under Grant No. NNX08AR22G and No. NNX12AR65G issued through the Planetary Science Division of the NASA Science Mission Directorate, the National Science Foundation under Grant No. AST-1238877, the University of Maryland, and Eotvos Lorand University (ELTE) and the Los Alamos National Laboratory. We thank the anonymous reviewer for useful discussions of systematic uncertainties and biases which strengthened the conclusions of this work. AM gratefully acknowledges support from DEL. AF acknowledges support from STFC research grant ST/L000709/1.

\nocite{topcat}
\bibliographystyle{mnras}
\bibliography{references}

\appendix

\label{lastpage}

\end{document}